\documentclass{menu99}    
\usepackage{epsfig}
\begin{document}
    \setlength{\baselineskip}{2.6ex}

\title{ How precisely can we determine the
$\pi$NN coupling constant from the isovector GMO sum rule?}
\author{B.~Loiseau \\
{\em LPNHE/LPTPE, Universit\'{e} P. \& M. Curie,
4 Place Jussieu, 75252 Paris, France} \\
T.E.O.~Ericson \\
{\em CERN, CH-1211 Geneva 23, Switzerland, and 
TSL, Box 533, S-75121 Uppsala, Sweden}
A.W.~Thomas \\
{\em CSSM, University of Adelaide, Adelaide 5005, Australia} }

\maketitle

\begin{abstract}
\setlength{\baselineskip}{2.6ex}
 The isovector GMO sum rule for zero energy forward
$\pi$N scattering is critically studied to obtain 
the charged $\pi N\! N$ coupling constant
using the precise  $\pi ^-$p and
$\pi^-$d scattering lengths deduced recently from pionic atom experiments.
This direct determination leads to
$g^2_c/4\pi =14.23\pm 0.09$~(statistic) $\pm 0.17$~(systematic) or
$f^2_c/4\pi = 0.0786(11)$. 
We obtain also accurate values for the $\pi$N scattering lengths
\end{abstract}

\section*{INTRODUCTION: ROBUST FORM OF THE GMO RELATION }

The analysis to  determine the $\pi $NN coupling
constant  should be clear and easily reproducible.
One should do a  detailed study for the
statistical and systematic errors.
The precise
determination
is an absolute statement, it could be
erroneous and it should be improvable.
In this perspective the Goldberger-Miyazawa-Oehme (GMO)
sum rule \cite{GOL55}  might be a good candidate.
It is a forward dispersion relation at zero energy for $\pi $N
scattering. It assumes  scattering amplitudes to be analytical
functions satisfying crossing symmetry.
At first isospin symmetry does not have to be assumed and it reads
(for more details see  e.g. \cite{HOE83})
with its numerical coefficients:
$ g_c^{2}/4\pi  = -4.50\ J^- +103.3\ [(a_{\pi ^- p}-a_{\pi ^+ p})/2]$,
where $J^-$ is in mb  the weighted integral,
$J^-=(1/4\pi ^2) \int_{0}^{\infty} (dk/\sqrt {k^2+m_{\pi}^2})
[\sigma ^{Total}_{\pi
^-p}(k) - \sigma^{Total}_{\pi ^+p}(k)]$
and $a_{\pi ^{\pm } p}$ are the $\pi^\pm p$ scattering lengths in units of
$m_{\pi}^{-1}$.
All ingredients are physical
observables but so far
the lack of precision in  $a_{\pi ^{\pm } p}$
(contribution of 2/3 to $g^2/4\pi$) led to
applications of
 the GMO
relation  as consistency check or constraint~\cite{ARN94b}.
The 1s width of the
 $\pi ^-p$ atom \cite{SCH99}  determines
$a_{\pi ^-p \to \pi ^0n} = -0.128(6)\ m_\pi^{-1}$
 and assuming isospin symmetry this
gives  $a^- = (a_{\pi ^- p}-a_{\pi ^+ p})/2$
 and $g_c^2/4\pi=14.2(4)$  using \cite{KOC85} $J^-$= -1.077(47) mb.
This is not accurate enough although improvements will come \cite{PSI99}.

We  here  report on a possible way to improve the precision on $g_c^2/4\pi$
\cite{ERI99}.
As $a_{\pi ^-p}$ is precisely known ($0.0883(8)\ m_{\pi}^{-1}$)
from energy shift in pionic hydrogen~\cite {CHA97} one can write:
\begin {equation}
 g_c^{2}/4\pi =
-4.50\ J^- +103.3\ a_{\pi ^- p}-103.3\ (\frac {a_{\pi ^- p}+a_{\pi ^+p}}{2}).
\label{eq:gmorobust}
\end{equation}
and using the above cited  $J^-$
(to be calculated later),
$g_c^{2}=4.85(22)+9.12(8)- 103.3 [(a_{\pi ^- p}+a_{\pi ^+p})/2]$.
This (not our final result)
shows that all the action is in the
 term $1/2(a_{\pi ^- p}+a_{\pi ^+p})$,which, assuming isospin symmetry,
 is 
 $a^+$. If this quantity is positive $g^2_c/4\pi$ is smaller than
14, if it is negative it is larger.
One way to determine  the small $a^+$ is to  
use the  accurate $\pi ^-d$
scattering length
$a_{\pi ^-d}=-0.0261(5)\ m_{\pi }^{-1},$
from the pionic deuterium 1s
energy level\cite{HAU98}.
To leading order this  is the coherent sum of
the $\pi ^-$ scattering lengths from the proton and 
neutron, which, assuming charge symmetry
(viz, $a_{\pi ^+p}=a_{\pi ^-n}$)
is  the term required in
our 'robust' relation~(\ref{eq:gmorobust})
 The strong cancellation between the two terms is then
done by the physics.
In order to
match the precision  using the width,
we only need a theoretical precision in the description of the deuteron
scattering length to about 30\%.

\section*{ZERO-ENERGY $\pi^-$-DEUTERON SCATTERING AND $a^+$}

In multiple scattering
theory of zero-energy s-wave
 pion scattering from  point-like nucleons and in the
fixed scattering-center approximation, the
leading contribution is\cite {ERI88}:
$a_{\pi ^-d}^{static}~= ~ S~+~D~ ...$ with
$ S=[(1+m_{\pi}/M)/(1+m_{\pi}/M_d)] (a_{\pi ^-p}+a_{\pi^-n}),$
 $M$ and $M_d$ being the nucleon and deuteron masses respectively.
The double scattering term $D$
is:
\begin {equation}
D~=~2~\frac {(1+m_{\pi}/M)^2}{(1+m_{\pi}/M_d)}~
[(\frac{a_{\pi ^-p}+a_{\pi^-n}}{2})^2-2(\frac {a_{\pi ^-p}-a_{\pi^-n}}{2})^2]
<1/r>,
\label{eq:D}
\end {equation}
and with our final
scattering lengths  $D=-0.0256\ m_{\pi }^{-1}$ 
quite close to 
$a_{\pi ^-d}$ experimental.

\parbox{8cm}{
\begin{center}
\epsfig{figure=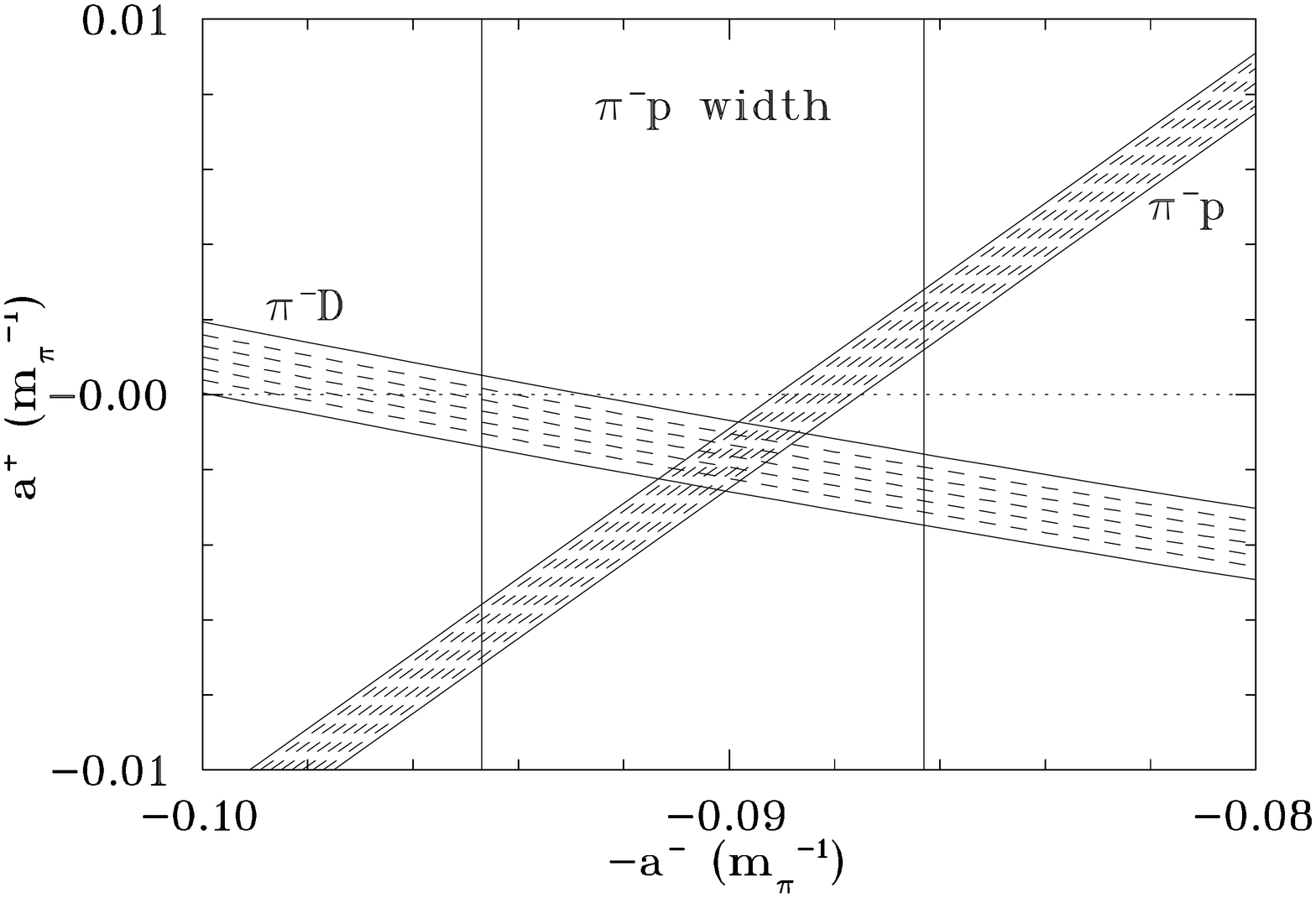,width=7cm,height=7cm,
angle=90}
\end{center}}
\parbox{7cm}{\vspace*{0.01cm}
\noindent
\parbox{5.2cm}
{\small \setlength{\baselineskip}{2.6ex} Fig.~1. Our graphical 
determination of the $\pi N$ scattering lengths 
in excellent agreement with the central values
of the experimental PSI group \cite {SCH99},
$a^+=-22(43)\cdot 10^{-4}m_{\pi}^{-1};a^-=905(42)\cdot 10^{-4}m_{\pi}^{-1}$.
}}

We shall here follow the recent theoretical  multiple scattering
investigation of
Baru and Kudryatsev (B-K) \cite{BAR97a}.
The comparison of typical contributions is listed in Table~1.
\begin{table}[htb]
\begin{center}
\caption[h]{ Typical contributions to
a$_{\pi d}$ in units of 10$^{-4}m_{\pi}^{-1}$, recall
a$_{\pi d}^{exp.}= -261(5)$\cite{HAU98}.}
{\small \begin{tabular}{|c|c|c|c|c|c|c|}
	\hline
	Contri & D & Fermi  & Absorption 
	 & s-p    & $(\pi ^- p,\gamma n)$  & Form   \\
          -butions &  & motion 1) & corr. \cite{THO80} 2) &interf. 3) 
         & double scatt. & factor 4) \\
	\hline
	Present  & -256(7)  &  61(7)   & -56(14)  &  small  & -2  & 17(9)   \\
	\hline
	B-K & -252 &  50 & X & -44 & X & 29(7)  \\
	\hline
	\hline
        Contri- & & Non-static  & Isospin  & Higher  & p-wave  & Virtual\\
butions && effects 5) & violation 6) & order 6) & double scatt. 6) & pion 6)\\ 
	\hline
	Present  & &10(6)  & 3.5  & 4(1)  & -3  & -7(2)     \\
	\hline
	B-K & &10 & 3.5 &6  & -3 & X   \\
	\hline
\end{tabular}}
\end{center}
\label{tab:contributions}
\end{table}
1){\it Fermi motion}: the nucleons have a momentum
distribution 
which produces an attractive
contribution, calculable to leading order from  $<p^2>$ of 
the nucleon momenta in the deuteron.
The uncertainty of 7 comes
from the D-state percentage in the deuteron, 
P$_D$ =4.3\% vs. 5.7\%
for the Machleidt1 \cite{MAC87} vs. the Paris \cite{LAC81} wave functions.
2) {\it Absorption correction}:
the absorption reaction $\pi^-$d $\to $nn,
using 3-body Faddeev approaches \cite {AFN74,MIZ77,THO80}
produces
a repulsive~($-$20\%) contribution (not included in B-K).
These studies were done carefully but a modern
reinvestigation of this term is highly desirable.
3) {\it ''s-p' interference}: a $-$15\% correction was obtained by B-K.
We find that it is a model dependent contribution due 
nearly entirely  
to the Born term
the contribution of which 
vanishes  exactly.
We have then not considered this contribution.
4){\it Form factor}:
this non-local effect enters
mainly via the dominant isovector $\pi $N s-wave
interaction,  closely linked to $\rho $ exchange.
It represents only a correction of about $-$10\%.
5) {\it Non-static effects}:  these produce only a rather small correction of
about 4\%. 
There are systematic cancellations 
between single and double scattering
as was first demonstrated
 by F\"aldt \cite
{FAL77}. It has been numerically investigated by B-K and
 we have adopted
their value, the error of 6 reflects a lack of
independent verification.
6) {\it Isospin violation, higher order terms, p-wave
double scattering,virtual pion scattering}: these corrections 
 are all small and controllable.
The isospin violation in the $\pi $N interaction 
comes in part from
the $\pi ^{\pm }-\pi ^0$ mass difference where an additional check 
comes from the chiral approach \cite {FET99}.
Based on this, we obtain
 the preliminary, though nearly final, values
$(a_{\pi ^-p}~+~a_{\pi ^-n})/2=(-17\pm
3 ({\rm statistic}) \pm 9 ({\rm systematic})) 10^{-4}m_{\pi }^{-1}$
and 
$(a_{\pi^-p}~-~a_{\pi^-n})/2=(900 \pm 12)~10^{-4}m_{\pi}^{-1}.$
Our values
represent a substantial improvement in accuracy as seen in
Fig.~1. 
The contribution of the scattering lengths to 
g$_c^2/4\pi$  has here a precision of about 1\%.

\section*{TOTAL CROSS SECTION INTEGRAL $J^-$}

  The cross section integral contributes only one third to the GMO relation. 
Total cross sections are 
inherently accurate and their contribution is calculated with accuracy, but
for the high energy region.
The possibility
of systematic effects in the difference must be considered, particularly
since Coulomb corrections have opposite sign for $\pi ^{\pm}$p.
The only
previous evaluation with a detailed discussion of errors is that by Koch
 \cite {KOC85}.
 Later evaluations given in
Table 2 find values within the errors, but the uncertainties are
not stated and analyzed.
\begin{table}[htb]
\begin{center}
\caption[h]{Some values of J$^-$ in mb, Ref. \cite {WOR92}
uses 2 different PWA: K-H \cite{KOC80} and their own, VPI.}
{\small \begin{tabular}{c|c|c|c|c|c|c}
 Ref.   &Koch               &  Workman {\it et al.}    & Workman {\it et al.}
        &Arndt {\it et al.}       & Gibbs {\it et al.}           & Present\\
         & 1985 \cite {KOC85}& 1992; K-H \cite {WOR92}&1992; VPI \cite {WOR92}
	  & 1995 \cite {ARN95}&1998 \cite {GIB98} & work\\
\hline
 J$^-$&-1.077(47)        &-1.056&-1.072&-1.05&-1.051&-1.095(31)\\
\end{tabular}}
\end{center}
\label{tab:J-values}
\end{table}
In view of obtaining a clear picture of the origin of
 uncertainties we have reexamined this problem in spite of the
 consensus.
We limit the discussion to the critical features.  The typical shape
of the integrands is shown in Fig.~2 up to 2 Gev/c.

\parbox{8cm}{
\begin{center}
\epsfig{figure=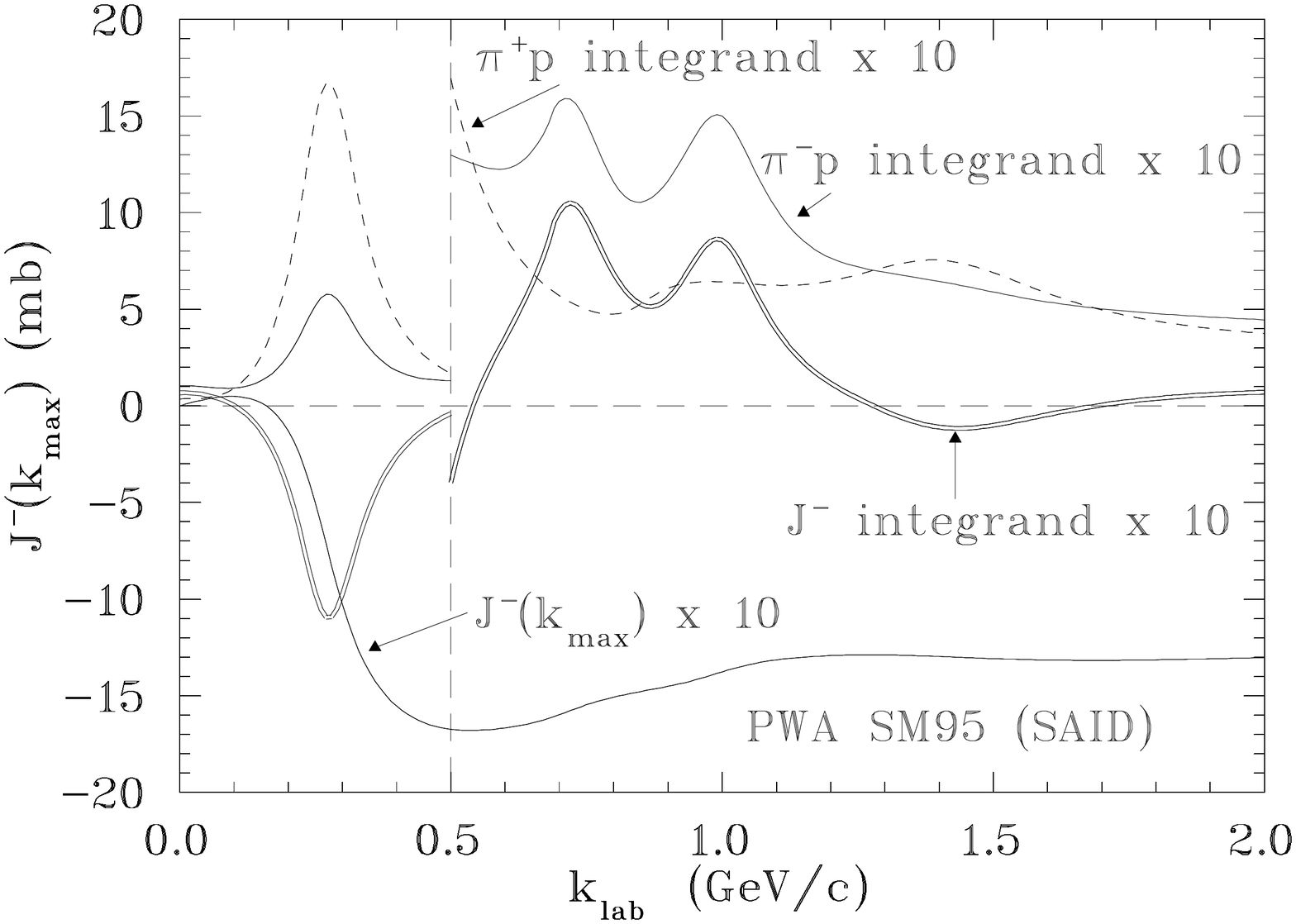,width=6cm,height=6cm,
angle=90}
\end{center}}
\parbox{7cm}{\vspace*{0.01cm}
\noindent
\parbox{5.2cm}
{\small \setlength{\baselineskip}{2.6ex} Fig.~2. The J$^-$ separate 
integrands  for $\pi^ {\pm}$p as well as 
their difference as function of k$_{lab}$ together with the cumulative
value of the integral J$^-(k_{lab})$ from the region
$0<k<k_{lab}$. The integrands are in units of mb GeV/c.}}

 There are
no total cross-section measurements  below 160 MeV/c, but the
accurately known $a_{\pi ^{\pm } p}$ give a strong constraint assuming
isospin symmetry. 
The s- and p-wave contributions  nearly cancels.
A tiny correction occurs,
since isospin is broken by the 3.3 MeV lower threshold for the $\pi ^0$n
channel below the physical $\pi ^-$p threshold.
The main contributions come from the region of the
$\Delta $ resonance and just above.
There are no strong cancellations in the difference between
$\pi^{\pm }$p
cross sections in that region and the cross sections have been
very carefully analyzed.
We
have first evaluated the  hadronic cross sections up to 2 GeV/c
based on the VPI phase-shift
solution \cite{ARN95b}. In doing so Coulomb
corrections and 
penetration factors have been  taken into
account in the adjustment to experimental data  even if the
treatment may not be optimal. It also allows for some isospin breaking,
since the $\Delta $ mass splitting is parameterized \cite{Pavan99}. In view
of the not so high accuracy we aim for, this
should be adequate.  Bugg \cite{BUG74,BUG99} has emphasized that
 in the $\pi ^+$p scattering the total cross sections are
systematically reduced at all energies by the Coulomb repulsion between
the particles and, conversely, enhanced in $\pi^-$p scattering.
 One must correct for this effect,
which gives a negative contribution to $J^-$.
Having made no correction for it at
higher energies  means that we will underestimate
the coupling constant somewhat.
In the region around 500 MeV/c there are long-standing problems with the
experimental total cross section data \cite{Pavan99}.
This uncertainty,
 larger than the Coulomb penetrability effects, should be
resolved.  So we have preferred to use the SM95 PWA
solution as the best guide.
  The real uncertainty in J$^-$ comes from the high energy region and is
linked to the relatively slow convergence of the integral.

\begin{table}[htb]
\begin{center}
\caption[h]{Contributions to J$^-$ in mb according to interval of
integration and to the total cross-section input. `Selected''
is for the world data \cite{PDG98} with statistical and systematic
errors $\le$ 1\%. Here the first given error is statistical and the
second one systematic.}
{\small \begin{tabular}{c|c|c|c|c}
k(GeV/c)  & 0 to 2                & 2 to 4.03           & 4.03 to 240
          & 240 to $\infty$       \\
\hline
Input     & SM95 \cite{ARN95b}    &Selected \cite{PDG98}& Selected \cite{PDG98}
          & Regge(94)\cite{PDG94} \\
\hline
 J$^-$(mb)&(-1.302$\pm 0.006$)(17)&(0.064$\pm 0.002$)(7)&(0.133$\pm 0.005$)(24)
          &(0.030)            \\
\hline
\hline
k(GeV/c)  & 0 to 2                & 240 to $\infty$    
          & 0 to $\infty$       & 0 to  $\infty$\\
\hline
Input     & Arndt98) \cite{ARN95b}    & Regge(98)\cite{PDG98} 
          & SM95+Regge94 & Arndt98+Regge98\\
\hline
 J$^-$(mb)&(-1.329$\pm 0.006$)(17)    &(0.018)(3)                
          &(-1.075$\pm 0.008)$(30)    &(-1.114$\pm 0.008)$(30)\\
\end{tabular}}
\end{center}
\label{tab:J-calculation}
\end{table}

We have evaluated the
different contributions (see Table~3)
with no other Coulomb and penetration corrections
than those introduced by the experimental authors above 2 GeV or by the
theoretical analysis below 2 GeV. We find, based on (integration of
hadronic cross section) the SM95 and Arndt
12/98 analysis below 2 GeV/c \cite{ARN95b}, and on the Regge pole PDG94
\cite{PDG94} and
PDG98 \cite{PDG98} extrapolation beyond 240 GeV/c, the values 
$J^-= (-1.075 \pm 0.008)(30) $ mb and
$(-1.114 \pm 0.008)(30)$ mb respectively.
 We have adopted the mean value
$J^-=(-1.095 \pm 0.008)(30)$ mb. 
In our calculation we have added a systematic uncertainty from Coulomb
penetration effect of $\pm 0.017$
from the region less than 2
GeV/c.

\section*{RESULTS AND CONCLUSIONS}
We have derived first new values for the  $\pi $N
scattering lengths from the $\pi ^-$d one:
$$a^+\simeq \frac {a_{\pi ^-p}+a_{\pi ^-n}}{2}=
(-17\pm 3)(9)) \cdot 10^{-4}m_{\pi}^{-1},
a^-\simeq  \frac {a_{\pi ^-p}-a_{\pi ^-n}}{2}=900(12)\cdot
10^{-4}m_{\pi}^{-1}.$$
Our second conclusion concerns the charged $\pi $NN coupling constant
using 
these
new accurate values
in (\ref{eq:gmorobust})  with 
$J^-=(-1.095 \pm 0.008)(30)$
and charge symmetry:  
\begin{equation}
g_c^{2}/4\pi  = (4.93 \pm 0.04)(14) + (9.12 \pm 0.08) + (0.18 \pm 0.03)(9)
=(14.23\pm 0.09)(17).\label{g^2final}
\end {equation}
The uncertainty 
comes mainly from $J^-$.
 This coupling constant which agrees quite well with the text book value,
 14.28(18) \cite{KOC80} is intermediate
between the low value deduced from the large data banks of NN and $\pi $N
scattering data \cite{Pavan99,Swart97}
and the high value from np charge exchange cross sections
\cite{RAH98}.
It is fully compatible with the latter, differing statistically by only
about one standard deviation.

\end{document}